\documentclass{article}
\usepackage{spconf,amsmath,graphicx}
\usepackage{extra_styles}
\usepackage[breaklinks,colorlinks]{hyperref}
\usepackage{url}
\usepackage{subfigure}

\usepackage{booktabs}
\usepackage{multirow}
\usepackage{amsmath}
\usepackage{amsfonts}

\title{FreGrad: lightweight and fast frequency-aware diffusion vocoder}
\name{Tan Dat Nguyen$^*$, Ji-Hoon Kim$^*$, Youngjoon Jang, Jaehun Kim, Joon Son Chung\thanks{$^*$These authors contributed equally to this work. This work was supported by the National Research Foundation of Korea grant funded by the Korean government (Ministry of Science and ICT, RS-2023-00212845) and the ITRC (Information Technology Research Center) support program (IITP-2024-RS-2023-00259991) supervised by the IITP (Institute for Information \& Communications Technology Planning \& Evaluation).}}
\address{Korea Advanced Institute of Science and Technology, South Korea}
\begin{document}
\ninept
\maketitle
\begin{abstract}
The goal of this paper is to generate realistic audio with a lightweight and fast diffusion-based vocoder, named FreGrad. Our framework consists of the following three key components: (1) We employ discrete wavelet transform that decomposes a complicated waveform into sub-band wavelets, which helps FreGrad to operate on a simple and concise feature space, (2) We design a frequency-aware dilated convolution that elevates frequency awareness, resulting in generating speech with accurate frequency information, and (3) We introduce a bag of tricks that boosts the generation quality of the proposed model. In our experiments, FreGrad achieves $3.7$ times faster training time and $2.2$ times faster inference speed compared to our baseline while reducing the model size by $0.6$ times (only $1.78$M parameters) without sacrificing the output quality. Audio samples are available at: \url{https://mm.kaist.ac.kr/projects/FreGrad}.
\end{abstract}
\section{Introduction}
\label{sec:intro}

Neural vocoder aims to generate audible waveforms from intermediate acoustic features (e.g. mel-spectrogram). It becomes an essential building block of numerous speech-related tasks
including singing voice synthesis~\cite{DBLP:conf/aaai/Liu00CZ22,DBLP:conf/kdd/RenTQLZL20}, voice conversion~\cite{DBLP:conf/icml/QianZCYH19,DBLP:conf/nips/ChoiLKLHL21}, and text-to-speech~\cite{DBLP:conf/icassp/ShenPWSJYCZWRSA18,DBLP:conf/icml/PopovVGSK21,DBLP:conf/nips/KimKKY20}.
Earlier neural vocoders~\cite{ISCA:2016:wavenet,DBLP:conf/iclr/MehriKGKJSCB17} are based on autoregressive (AR) architecture, demonstrating the ability to produce highly natural speech.
However, their intrinsic architecture requires a substantial number of sequential operations, leading to an extremely slow inference speed.
Numerous efforts in speeding up the inference process have been made on non-AR architecture based on flow~\cite{ICASSP:2019:waveglow,ICML:2020:WaveFlow}, generative adversarial networks~\cite{DBLP:conf/nips/KumarKBGTSBBC19,DBLP:conf/iclr/EngelACGDR19,ICASSP:2020:ParallelWaveGAN}, and signal processing \cite{DBLP:journals/taslp/JuvelaBTA19,DBLP:conf/icassp/KanekoTKS22}. While such approaches have accelerated the inference speed, they frequently produce lower quality waveforms compared to AR methods.
Among non-AR vocoders, diffusion-based vocoders have recently attracted increasing attention due to its promising generation quality~\cite{ICLR:2021:DiffWave, DBLP:conf/iclr/ChenZZWNC21, IJCAI:2022:FastDiff, ICLR:2022:PriorGrad, DBLP:conf/iclr/Lam00022,DBLP:conf/interspeech/KoizumiZYCB22,10095749}.
Despite its high-quality synthetic speech, diffusion-based vocoder suffers from slow training convergence speed, inefficient inference process, and high computation cost.
These factors hinder the utilization of diffusion-based vocoders in low-resource devices and their application in real-world scenarios.
While many works~\cite{IJCAI:2022:FastDiff,DBLP:conf/iclr/Lam00022,DBLP:conf/icassp/ChenTWPMHZ22} have tried to minimize training and inference times, there still remains a limited exploration to reduce computational costs.
\begin{figure}
    \centering
    \includegraphics[width=0.45\textwidth]{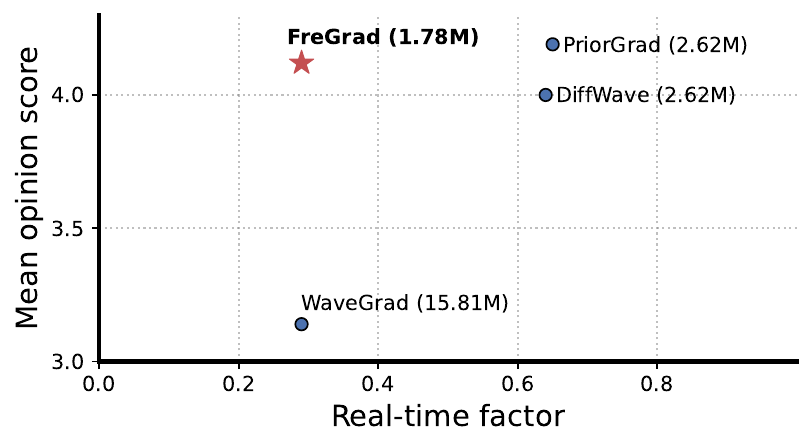}
    \caption{FreGrad successfully reduces both real-time factor and the number of parameters while maintaining the synthetic quality.}
    \label{fig:model_comparision}
\end{figure}

\begin{figure*}[t]
    \centering
    \includegraphics[width=.9\linewidth]{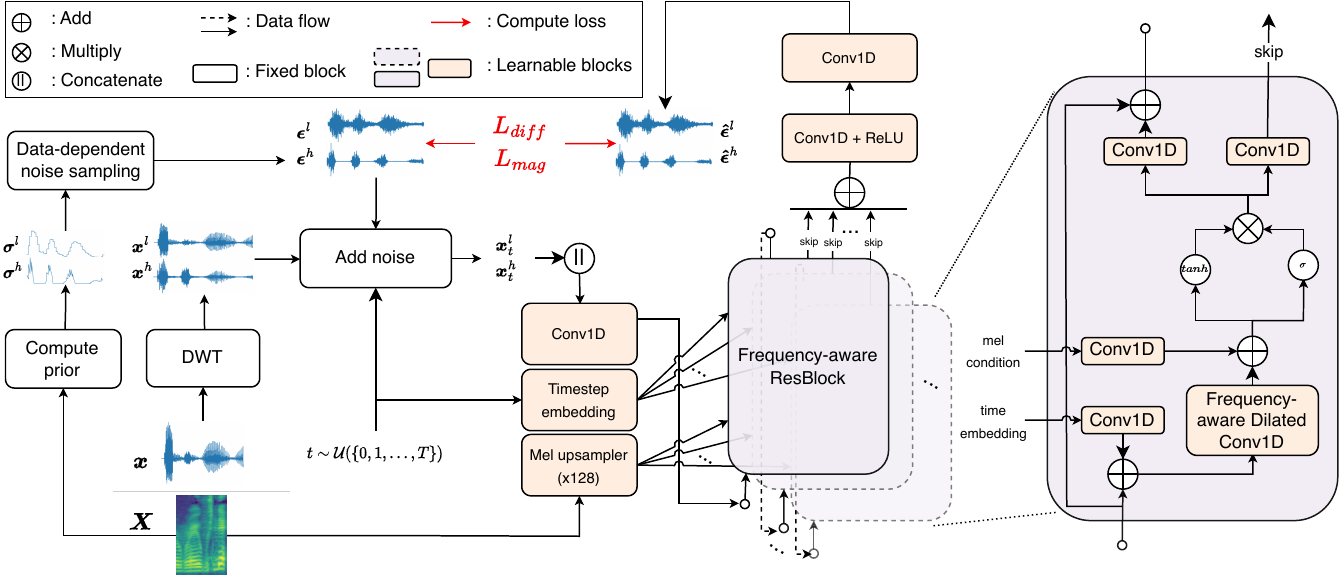}
    \caption{Training procedure and model architecture of FreGrad. We compute wavelet features $\{\boldsymbol{x}^l$, $\boldsymbol{x}^h\}$ and prior distributions $\{\boldsymbol{\sigma}^l$, $\boldsymbol{\sigma}^h\}$ from waveform $\boldsymbol{x}$ and mel-spectrogram $\boldsymbol{X}$, respectively. At timestep $t$, noises $\{\boldsymbol{\epsilon}^l$, $\boldsymbol{\epsilon}^h\}$ are added to each wavelet feature. Given mel-spectrogram and timestep embedding, FreGrad approximates the noises $\{\boldsymbol{\hat\epsilon}^l$, $\boldsymbol{\hat\epsilon}^h\}$. The training objective is a weighted sum of $\mathcal{L}_{diff}$ and $\mathcal{L}_{mag}$ between ground truth and the predicted noise.}
    \vspace{-1mm}
    \label{fig:Training procedure}
\end{figure*} 

To address the aforementioned problems at once, in this paper, we propose a novel diffusion-based vocoder called FreGrad, which achieves both low memory consumption and fast processing speed while maintaining the quality of the synthesized audio.
The key to our idea is to decompose the complicated waveform into two simple frequency sub-band sequences (i.e. wavelet features), which allow our model to avoid heavy computation.
To this end, we utilize discrete wavelet transform (DWT) that converts a complex waveform into two frequency-sparse and dimension-reduced wavelet features without a loss of information~\cite{DBLP:books/siam/92/D1992,ISCA:Fre-GAN:2021}.
FreGrad successfully reduces both the model parameters and denoise processing time by a significant margin.
In addition, we introduce a new building block, named frequency-aware dilated convolution (Freq-DConv), which enhances the output quality.
By incorporating DWT into the dilated convolutional layer, we provide the inductive bias of frequency information to the module, and thereby the model can learn accurate spectral distributions which serves as a key to realistic audio synthesis.
To further enhance the quality, we design a prior distribution for each wavelet feature, incorporate noise transformation that replaces the sub-optimal noise schedule, and leverage a multi-resolution magnitude loss function that gives frequency-aware feedback.

In the experimental results, we demonstrate the effectiveness of FreGrad with extensive metrics.
FreGrad demonstrates a notable enhancement in boosting model efficiency while keeping the generation quality.
As shown in \Tref{tab:objective_results}, FreGrad boosts inference time by 2.2 times and reduces the model size by 0.6 times with mean opinion score (MOS) comparable to existing works. 

\section{Backgrounds}
\label{sec:background}
The denoising diffusion probabilistic model is a latent variable model that learns a data distribution by denoising a noisy signal~\cite{DBLP:conf/nips/HoJA20}.
The \textit{forward} process $q(\cdot)$ diffuses data samples through Gaussian transitions parameterized
with a Markov process:
\begin{equation}
q(\boldsymbol{x}_t \vert \boldsymbol{x}_{t-1}) = \mathcal{N}(\boldsymbol{x}_t ; \sqrt{1 - \beta_t} \boldsymbol{x}_{t-1}, \beta_t \mathbf{I}),
\end{equation}
where $\beta_t \in \{\beta_1, \ldots, \beta_T \}$ is the predefined noise schedule, $T$ is the total number of timesteps, and $\boldsymbol{x}_0$ is the ground truth sample.
This function allows sampling $\boldsymbol{x}_t$ from $\boldsymbol{x}_0$, which can be formulated as:
\begin{equation}
    \boldsymbol{x}_t = \sqrt{\gamma_t} \boldsymbol{x}_0 + \sqrt{1 - \gamma_t} \boldsymbol \epsilon,
    \label{eq:add_noise}
\end{equation}
where $\gamma_t = \prod_{i=1}^t ( 1-\beta_i)$ and $\boldsymbol\epsilon \sim \mathcal{N}(\boldsymbol{0}, \mathbf{I})$.

With a sufficiently large $T$, the distribution of $\boldsymbol{x}_T$ approximates an Isotropic Gaussian distribution.
Consequently, we can obtain a sample in ground truth distribution by tracing the exact \textit{reverse} process $p(\boldsymbol{x}_{t-1} \vert \boldsymbol{x}_t)$ from an initial point $\boldsymbol{x}_T \sim \mathcal{N}(\boldsymbol{0}, \mathbf{I})$.
Since $p(\boldsymbol{x}_{t-1} \vert \boldsymbol{x}_t)$ depends on the entire data distribution, we approximate it with a neural network $p_{\boldsymbol\theta} (\boldsymbol{x}_{t-1} \vert \boldsymbol{x}_t)$ which is defined as $\mathcal{N}(\boldsymbol{x}_{t-1}; \mu_{\boldsymbol\theta}(\boldsymbol{x}_t, t), \sigma^2_{\boldsymbol\theta}(\boldsymbol{x}_t, t))$. 
As shown in \cite{DBLP:conf/nips/HoJA20}, the variance $ \sigma^2_{\boldsymbol\theta}(\cdot)$ can be represented as $\frac{1 - \gamma_{t-1}}{1 - \gamma_{t}}\beta_t$, and mean $\mu_{\boldsymbol\theta}(\cdot)$ is given by:
\begin{equation}
    \mu_{\boldsymbol{\theta}}(\boldsymbol{x}_t, t) = \frac{1}{\sqrt{1-\beta_t}}\left(\boldsymbol{x}_t - \frac{\beta_t}{\sqrt{1 - \gamma_t}} \epsilon_{\boldsymbol\theta}(\boldsymbol{x}_t, t)\right),
\end{equation}
where $\epsilon_{\boldsymbol\theta}(\cdot)$ is a neural network that learns to predict the noise.

In practice, the training objective for $\epsilon_{\boldsymbol\theta}(\cdot)$ is simplified to minimize
$\mathbb{E}_{t, \boldsymbol{x}_t, \epsilon} \left[\Vert \boldsymbol \epsilon - \epsilon_{\boldsymbol\theta}(\boldsymbol{x}_t, t) \Vert_2^2 \right]
$. PriorGrad \cite{ICLR:2022:PriorGrad} extends the idea by starting the sampling procedure from the prior distribution $\mathcal{N}(\mathbf{0}, \mathbf{\Sigma})$.
Here, $\mathbf{\Sigma}$  is a diagonal matrix $ diag\left[(\sigma_0^2, \sigma_1^2, \ldots, \sigma_N^2)\right]$, where $\sigma_i^2$ is the $i^{th}$ normalized frame-level energy of mel-spectrogram with length $N$. Accordingly, the loss function for $\epsilon_{\boldsymbol\theta}(\cdot)$ is modified as:
\begin{equation}
    \mathcal{L}_{diff} = \mathbb{E}_{t, \boldsymbol x_t, \boldsymbol\epsilon, \boldsymbol c}\left[\Vert \boldsymbol{\epsilon} - \epsilon_\theta (\boldsymbol x_t, t, \boldsymbol X) \Vert_{\boldsymbol\Sigma^{-1}}^2\right],
    \label{eq:priorgrad_loss}
\end{equation}
where $\Vert \boldsymbol{x} \Vert_{\boldsymbol\Sigma^{-1}}^2 = \boldsymbol{x}^\top \boldsymbol{\Sigma}^{-1} \boldsymbol{x}$ and $\boldsymbol X$ is a mel-spectrogram.

\section{FreGrad}
\label{sec:method}
The network architecture of FreGrad is rooted in DiffWave~\cite{ICLR:2021:DiffWave} which is a widely used backbone network for diffusion-based vocoders~\cite{ICLR:2022:PriorGrad,10095749}. 
However, our method is distinct in that it operates on a concise wavelet feature space and replaces the existing dilated convolution with the proposed Freq-DConv to reproduce accurate spectral distributions.

\subsection{Wavelet Features Denoising}
To avoid complex computation, we employ DWT before \textit{forward} process. 
DWT downsamples the target dimension audio $\boldsymbol{x}_0 \in \mathbb{R}^{L}$ into two wavelet features $\{\boldsymbol{x}^{l}_0, \boldsymbol{x}^{h}_0\}\subset\mathbb{R}^{\frac{L}{2}}$, each of which represents low- and high-frequency components.
As demonstrated in the previous works~\cite{ISCA:Fre-GAN:2021,ICASSP:2022:FreGAN2}, the function can deconstruct a non-stationary signal without information loss due to its biorthogonal property.

FreGrad operates on simple wavelet features. 
At each training step, the wavelet features $\boldsymbol{x}^{l}_0$ and $\boldsymbol{x}^{h}_0$ are diffused into noisy features at timestep $t$ with distinct noise $\boldsymbol{\epsilon}^{l}$ and $\boldsymbol{\epsilon}^{h}$, and each noise is simultaneously approximated by a neural network $\epsilon_{\boldsymbol\theta}(\cdot)$.
In \textit{reverse} process, FreGrad simply generates denoised wavelet features, $\{\boldsymbol{\hat{x}}^{l}_0$, $\boldsymbol{\hat{x}}^{h}_0\} \subset \mathbb{R}^{\frac{L}{2}}$, which are finally converted into the target dimensional waveform $\boldsymbol{\hat{x}}_0\in\mathbb{R}^L$ by inverse DWT (iDWT):
\begin{equation}
    \boldsymbol{\hat{x}}_0 =  \Phi^{-1}(\boldsymbol{\hat{x}}^{l}_0, \boldsymbol{\hat{x}}^{h}_0),
\end{equation}
where  $\Phi^{-1}(\cdot)$ denotes the iDWT function. 

Note that FreGrad generates speech with smaller computations due to the decomposition of complex waveforms. In addition, the model maintains its synthetic quality, as iDWT guarantees a lossless reconstruction of a waveform from wavelet features~\cite{ICASSP:2022:FreGAN2,reichel2001integer}.
In our experiments, we adopt Haar wavelet~\cite{haar1909theorie}.

\subsection{Frequency-aware Dilated Convolution}
\begin{figure}[t]
  \begin{center}
    \includegraphics[width=0.17\textwidth, angle=90]{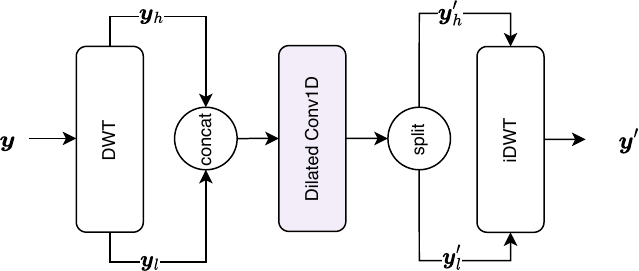}
  \end{center}
  \vspace{-.5mm}
  \caption{Frequency-aware dilated convolution.}
  \label{fig:freq_dilated_conv}
  \vspace{-3.5mm}
\end{figure}

Since audio is a complicated mixture of various frequencies~\cite{ISCA:Fre-GAN:2021}, it is important to reconstruct accurate frequency distributions for natural audio synthesis.
To enhance the synthetic quality, we propose Freq-DConv which deliberately guides the model to pay attention to the frequency information.
As illustrated in \Fref{fig:freq_dilated_conv}, we adopt DWT to decompose the hidden signal $\boldsymbol y \in \mathbb{R}^{\frac{L}{2}\times D}$ into two sub-bands $\{\boldsymbol{y}_l$, $\boldsymbol{y}_h\} \subset \mathbb{R}^{\frac{L}{4}\times D}$ with hidden dimension $D$.
The sub-bands are channel-wise concatenated, and the following dilated convolution $\mathbf{f}(\cdot)$ extracts a frequency-aware feature $\boldsymbol{y}_{hidden} \in \mathbb{R}^{\frac{L}{4}\times 2 D}$: 
\begin{equation}
    \boldsymbol{y}_{hidden} = \mathbf{f}(\mathtt{cat}(\boldsymbol{y}_l, \boldsymbol{y}_h)),
\end{equation}
where $\mathtt{cat}$ denotes concatenation operation.
The extracted feature $\boldsymbol{y}_{hidden}$ is then bisected into $\{\boldsymbol{y}'_l,\boldsymbol{y}'_h\}\subset \mathbb{R}^{\frac{L}{4}\times D}$ along channel dimension, and finally iDWT converts the abstract features into single hidden representation to match the length with input feature $\boldsymbol{y}$: 
\begin{equation}
    \boldsymbol{y}' = \Phi^{-1}(\boldsymbol{y}'_l,\boldsymbol{y}'_h),
\end{equation}
where $\boldsymbol{y}'\in\mathbb{R}^{\frac{L}{2}\times D}$ represents the output of the Freq-DConv.
As depicted in \Fref{fig:Training procedure}, we embed the Freq-DConv into every ResBlock.

The purpose of decomposing the hidden signal before the dilated convolution is to increase the receptive field along the time axis without changing the kernel size.
As a result of DWT, each wavelet feature has a reduced temporal dimension while preserving all temporal correlations.
This helps each convolution layer to possess a larger receptive field along the time dimension even with the same kernel size.
Furthermore, low- and high-frequency sub-bands of each hidden feature can be explored separately. 
As a result, we can provide an inductive bias of frequency information to the model, which facilitates the generation of frequency-consistent waveform.
We verify the effectiveness of Freq-DConv in Sec.~\ref{subsec:ablation}.
 
\subsection{Bag of Tricks for Quality}

\newpara{Prior distribution.}
\label{ssec:prior computation}
As demonstrated in previous works~\cite{ICLR:2022:PriorGrad,DBLP:conf/interspeech/KoizumiZYCB22}, a spectrogram-based prior distribution can significantly enhance the waveform denoising performance even with fewer sampling steps.
Building upon this, we design a prior distribution for each wavelet sequence based on the mel-spectrogram. 
Since each sub-band sequence contains specific low- or high-frequency information, we use separate prior distribution for each wavelet feature. 
Specifically, we divide the mel-spectrogram into two segments along the frequency dimension and adopt the technique proposed in \cite{ICLR:2022:PriorGrad} to obtain separate prior distributions $\{\boldsymbol{\sigma}^{l},\boldsymbol\sigma^{h}\}$ from each segment.

\begin{figure}[t]
\begin{minipage}[b]{.45\linewidth}
  \centerline{\includegraphics[width=4.5cm]{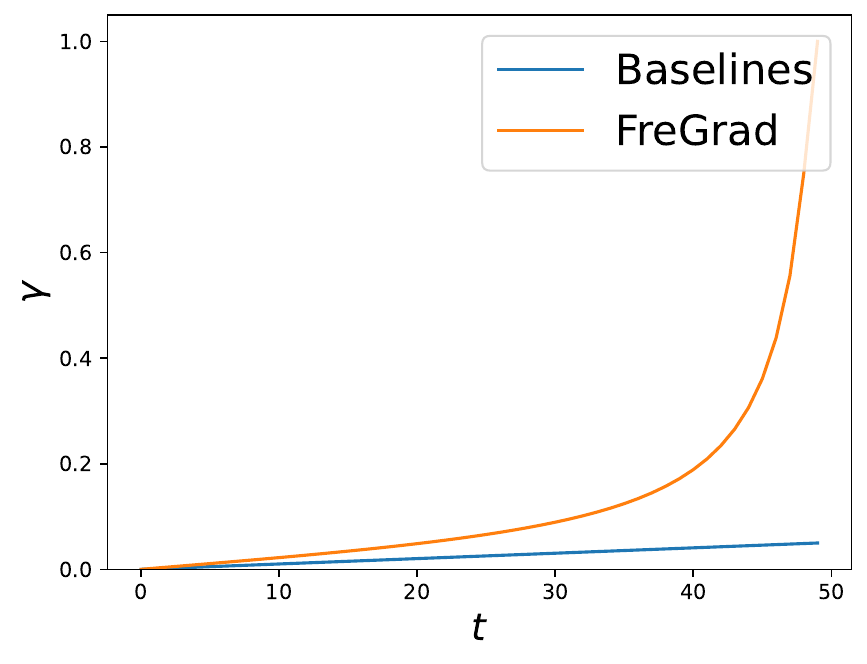}}
  \centerline{(a) Noise level $\boldsymbol \gamma$}\medskip
\end{minipage}
\hfill
\begin{minipage}[b]{0.45\linewidth}
  \centerline{\includegraphics[width=4.5cm]{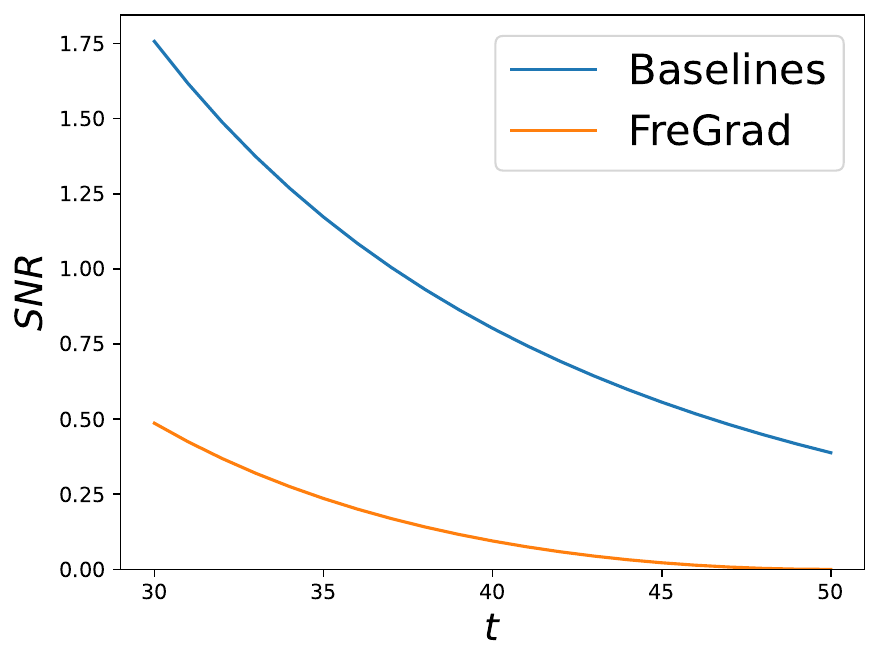}}
  \centerline{(b)  $\text{SNR}$}\medskip
\end{minipage}
\vspace{-2mm}
\caption{Noise level and log SNR through timesteps. ``Baselines" refer to the work of \cite{ICLR:2021:DiffWave,DBLP:conf/iclr/ChenZZWNC21,ICLR:2022:PriorGrad} which use the same linear beta schedule $\boldsymbol\beta$ ranging from $0.0001$ to $0.05$ for 50 diffusion steps.}
\label{fig:noise_level}
\vspace{-2mm}
\end{figure}

\newpara{Noise schedule transformation.}
As discussed in \cite{DBLP:journals/corr/abs-2301-11093,lin2023common}, signal-to-noise ratio (SNR) should ideally be zero at the final timestep $T$ of \textit{forward} process.
However, noise schedules adopted in previous works \cite{ICLR:2021:DiffWave,DBLP:conf/iclr/ChenZZWNC21,ICLR:2022:PriorGrad} fail to reach SNR near zero at the final step, as shown in \Fref{fig:noise_level}.
To achieve a zero SNR at the final step, we adopt the proposed algorithm in \cite{lin2023common}, which can be formulated as follows:
\begin{equation}
    \sqrt{\boldsymbol{\gamma}}_{new} = \frac{\sqrt{\gamma}_0}{\sqrt{\gamma}_0 - \sqrt{\gamma}_T + \tau} (\sqrt{\boldsymbol{\gamma}} - \sqrt{\gamma}_T + \tau),
    \label{eq:shift_function}
\end{equation}
where $\tau$ helps to avoid division by zero in sampling process.

\newpara{Loss function.}
A common training objective of diffusion vocoder is to minimize the L2 norm between predicted and ground truth noise, which lacks explicit feedbacks in the frequency aspect.
To give a frequency-aware feedback to the model, we add multi-resolution short-time Fourier transform (STFT) magnitude loss ($\mathcal{L}_{mag}$). 
Different from the previous works~\cite{ICASSP:2020:ParallelWaveGAN, DBLP:conf/icassp/ChenTWPMHZ22}, FreGrad only uses magnitude part since we empirically find that integrating \textit{spectral convergence loss} downgrades the output quality. Let $M$ be the number of STFT losses, then $\mathcal{L}_{mag}$ can be represented as:
\begin{equation}
    \mathcal{L}_{mag}= \frac{1}{M} \sum_{i=1}^M \mathcal{L}_{mag}^{(i)},
\end{equation}
where $\mathcal{L}_{mag}^{(i)}$ is STFT magnitude loss from $i^{th}$ analysis settings~\cite{ICASSP:2020:ParallelWaveGAN}. We separately apply the diffusion loss to low- and high-frequency sub-bands, and the final training objective is defined as:
\begin{equation}
    \mathcal{L}_{final} = \sum_{i \in \{l, h\}} \left[ \mathcal{L}_{diff} (\boldsymbol\epsilon^{i}, \boldsymbol{\hat{\epsilon}}^i) + \lambda \mathcal{L}_{mag} (\boldsymbol{\epsilon}^i , \boldsymbol{\hat{\epsilon}}^i) \right],
    \label{eq:loss_function}
\end{equation}
where $\boldsymbol{\hat{\epsilon}}$ refers to an estimated noise.

\begin{table*}
    \centering
        \caption{Evaluation results. The MOS results are presented with 95\% confidence intervals. $\uparrow$ means higher is better, $\downarrow$ denotes lower is better.}
        \label{tab:objective_results}
        \resizebox{\textwidth}{!}{
    \begin{tabular}{lcccccccc}
    \toprule
    \textbf{Model} & MOS $\uparrow$ & MAE $\downarrow$  & MR-STFT $\downarrow$ & MCD$_{13}$ $\downarrow$     & RMSE$_{f_0}$  $\downarrow$ & \#params $\downarrow$ &RTF on CPU $\downarrow$ & RTF on GPU $\downarrow$ \\ \midrule
    Ground truth & $4.74 \pm 0.06$ & $-$ & $-$ & $-$ & $-$ & $-$ & $-$ & $-$ \\ \midrule
    WaveGrad  & $3.14 \pm 0.09$ & $0.59$ & $1.39 $ & $3.06 $ & $39.97 $  & $15.81$M      & $\mathbf{11.58}$ & $\boldsymbol{0.29}$  \\ 
    DiffWave  & $4.00 \pm 0.10$& $0.56$ & $1.18 $  &  $3.20 $ & $40.10 $ & ~~$2.62$M      & $29.99$ & $0.64$ \\ 
    PriorGrad & $\mathbf{4.19 \pm 0.10}$ & $0.47 $ & $1.14 $ & $2.22 $ & $40.42 $ & ~~$2.62$M      &  $29.20$ & $0.65$ \\ \midrule
    
    \textbf{FreGrad} & ${4.12 \pm 0.11}$   & $\mathbf{0.45}$ & $\mathbf{1.12}$  & $ \mathbf{2.19}$ & $\mathbf{38.73}$ & ~~$\mathbf{1.78}$M      & ${11.95}$ & $\mathbf{0.29}$ \\ \bottomrule
    \end{tabular}
    }
\end{table*}

\begin{figure*}[t]
\begin{minipage}[b]{.33\linewidth}
  \centerline{\includegraphics[width=5cm]{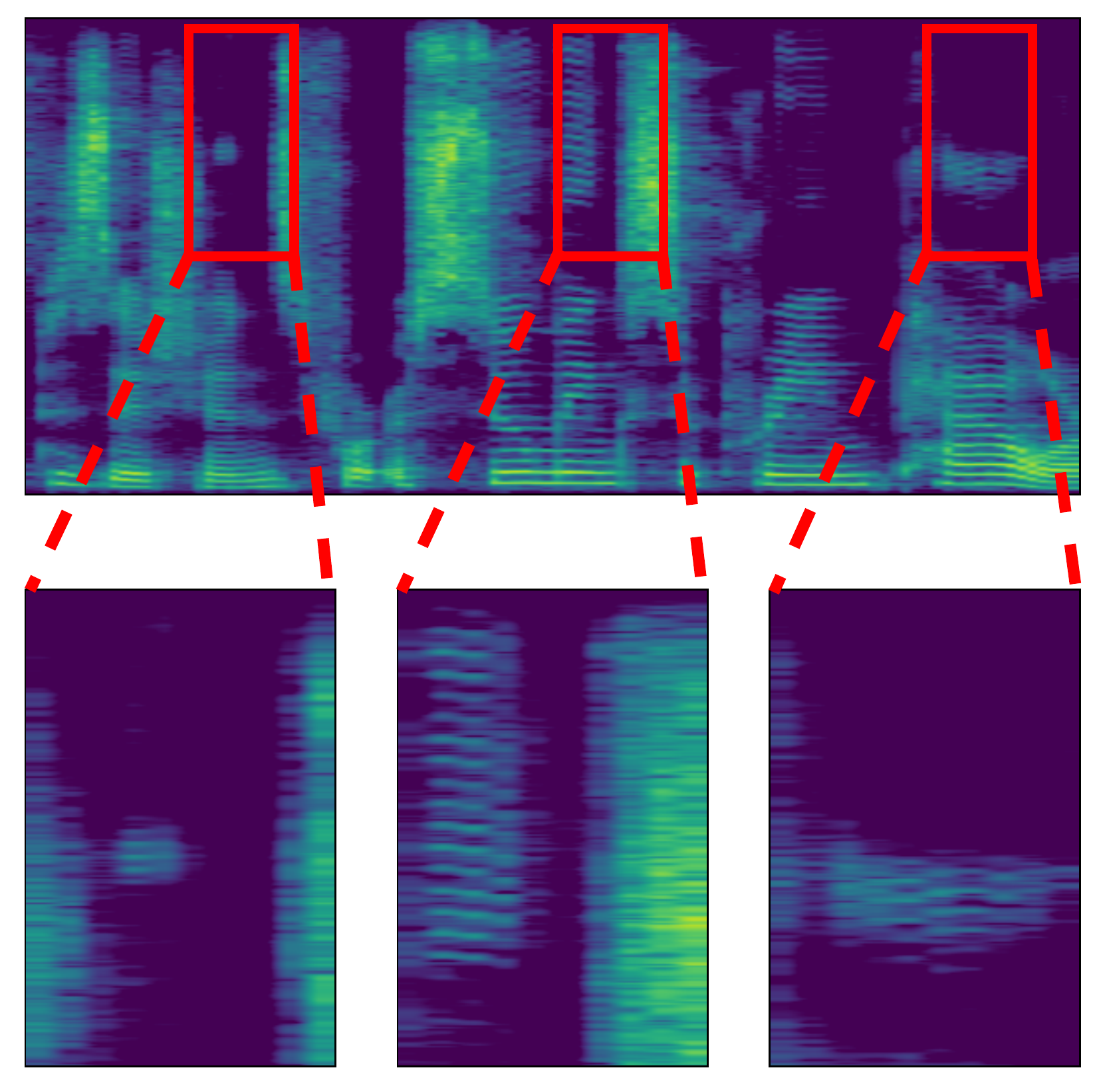}}
  \centerline{(a) Ground truth}\medskip
\end{minipage}
\hfill
\begin{minipage}[b]{0.33\linewidth}
  \centerline{\includegraphics[width=5cm]{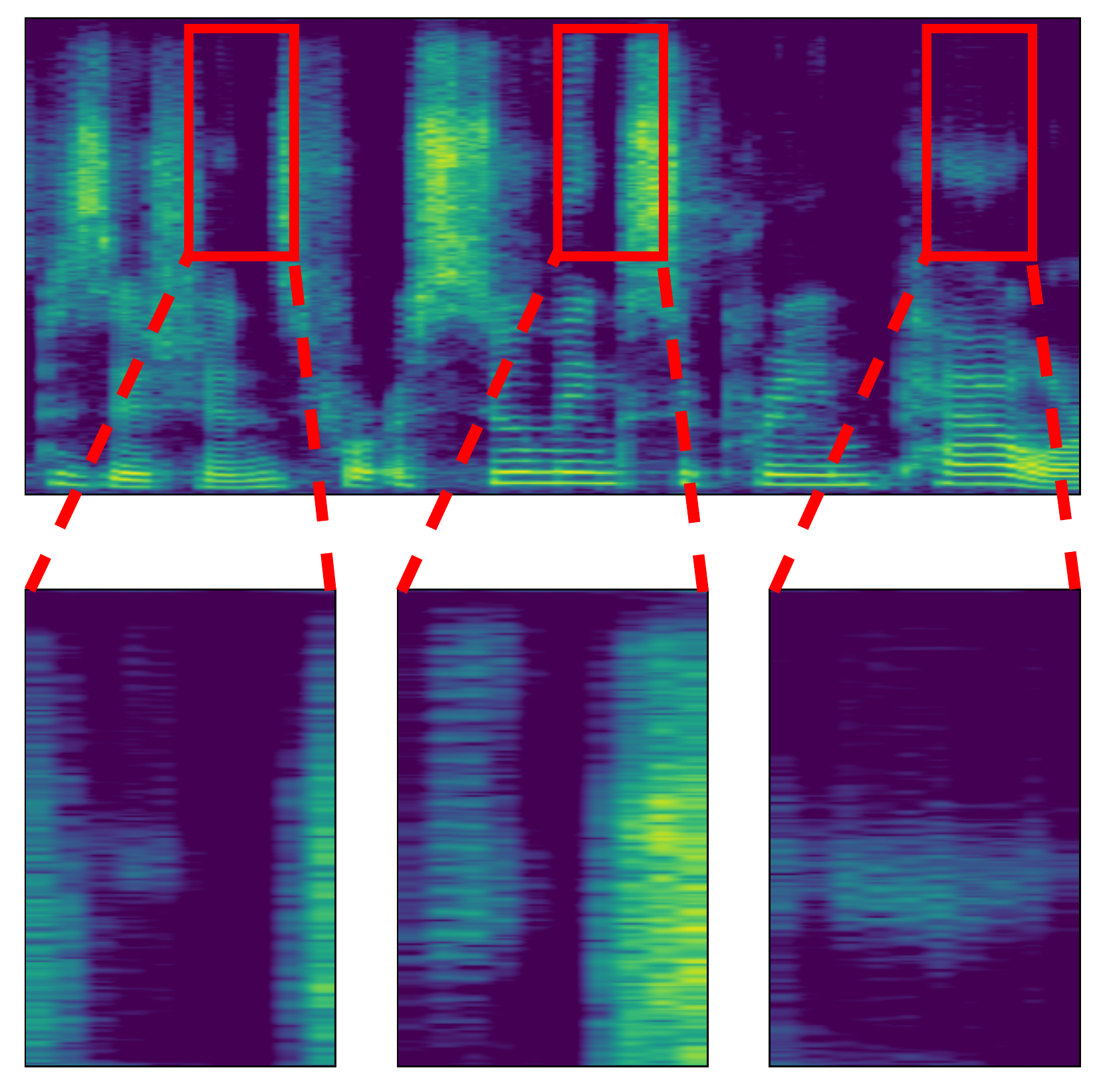}}
  \centerline{(b) FreGrad}\medskip
\end{minipage}
\hfill
\begin{minipage}[b]{0.33\linewidth}
  \centerline{\includegraphics[width=5cm]{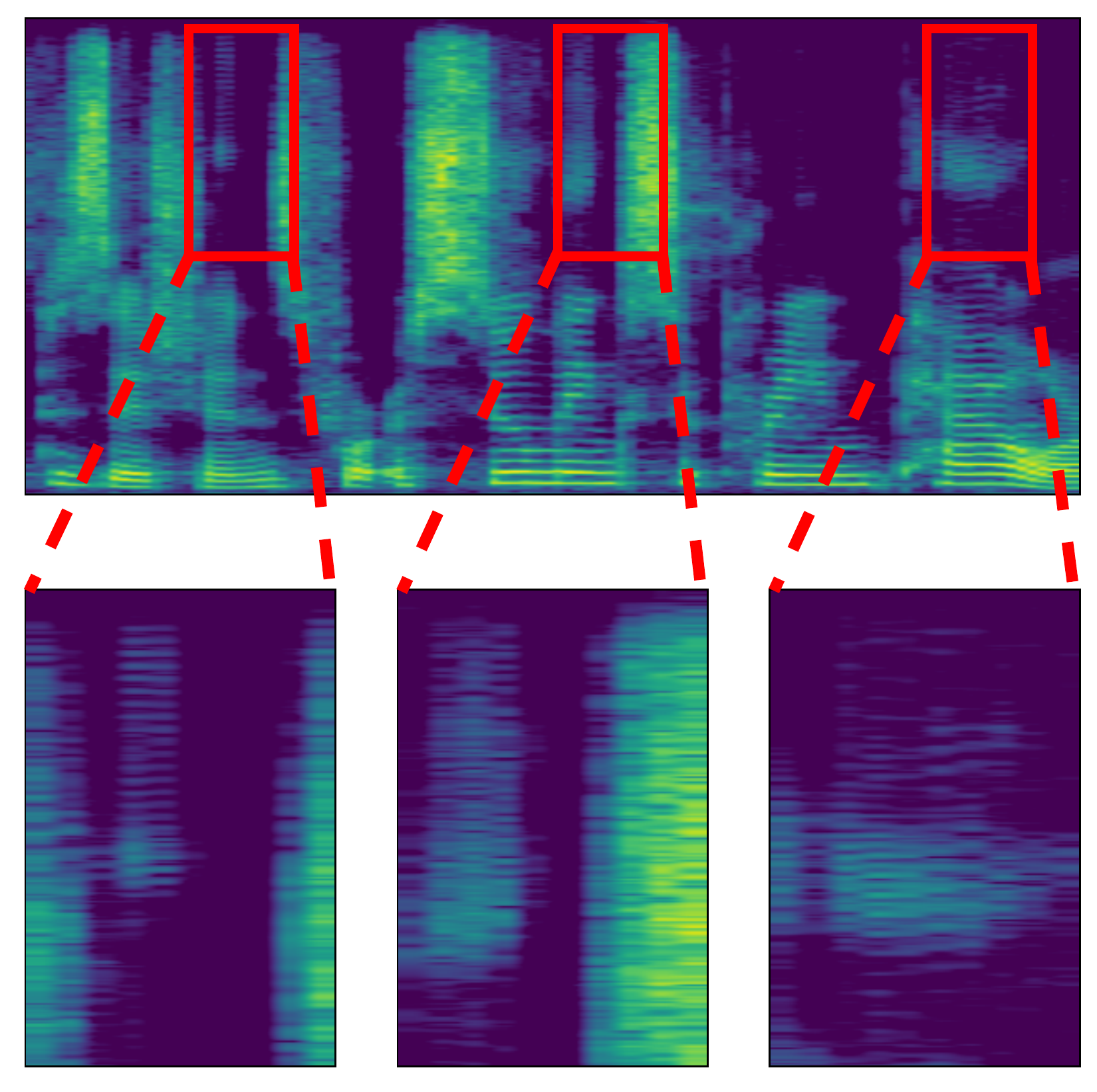}}
  \centerline{(c) PriorGrad}\medskip
\end{minipage}
\vspace{-8mm}
\caption{Spectrogram analysis on FreGrad and PriorGrad. While PriorGrad suffers from over-smoothed results, FreGrad reproduces detailed spectral correlation, especially in red boxes.}
\vspace{-3mm}
\label{fig:mel_comparison}
\end{figure*}

\section{Experiments}

\subsection{Training Setup}
We conduct experiments on a single English speaker LJSpeech\footnote{https://keithito.com/LJ-Speech-Dataset} which contains 13,100 samples. We use 13,000 random samples for training and 100 remaining samples for testing. 
Mel-spectrograms are computed from the ground truth audio with 80 mel filterbanks, 1,024 FFT points ranging from 80Hz to 8,000Hz, and hop length of 256. FreGrad is compared against the best performing publicly available diffusion vocoders: WaveGrad\footnote{https://github.com/lmnt-com/wavegrad}, DiffWave\footnote{https://github.com/lmnt-com/diffwave}, and PriorGrad\footnote{https://github.com/microsoft/NeuralSpeech}. For fair comparison, all the models are trained until 1M steps, and all the audios are generated through 50 diffusion steps which is the default setting in DiffWave~\cite{ICLR:2021:DiffWave} and PriorGrad~\cite{ICLR:2022:PriorGrad}.

FreGrad consists of $30$ frequency-aware residual blocks with a dilation cycle length of $7$ and a hidden dimension of $32$. We follow the implementation of DiffWave~\cite{ICLR:2021:DiffWave} for timestep embedding and mel upsampler but reduce the upsampling rate by half because the temporal length is halved by DWT.
For $\mathcal{L}_{mag}$, we set $M=3$ with FFT size of $[512, 1024, 2048]$ and window size of $[240, 600, 1200]$. We choose $\tau = 0.0001$ and $\lambda=0.1$ for \Eref{eq:shift_function} and \Eref{eq:loss_function}, respectively. We utilize Adam optimizer with $\beta_1 = 0.9$, $\beta_2=0.999$, fixed learning rate of $0.0002$, and batch size of 16.

\subsection{Audio Quality and Sampling Speed}
We verify the effectiveness of FreGrad on various metrics. To evaluate the audio quality, we obtain mel-cepstral distortion (MCD$_{13}$) and 5-scale MOS where 25 subjects rate the naturalness of 50 audio samples. In addition, we compute mean absolute error (MAE), $f0$ root mean square error (RMSE$_{f0}$), and multi-resolution STFT error (MR-STFT) between generated and ground truth audio. To compare the model efficiency, we calculate the number of model parameters (\#params) and real-time factor (RTF) which is measured on AMD EPYC 7452 CPU and a single GeForce RTX 3080 GPU. Except for MOS, all the metrics are obtained from 100 audio samples.

As demonstrated in~\Tref{tab:objective_results}, FreGrad highly reduces not only the number of model parameters but also inference speed on both CPU and GPU. In addition, FreGrad achieves the best results in all the quality evaluation metrics except for MOS. 
Given humans' heightened sensitivity to low-frequency sounds, we hypothesize that the MOS degradation in FreGrad results from low-frequency distribution.
However, in perspective of the entire spectrum of frequencies, FreGrad consistently demonstrates superior performance compared to existing methods, as confirmed by the MAE, MR-STFT, MCD$_{13}$, and RMSE$_{f0}$.
The mel-spectrogram visualization analysis (\Fref{fig:mel_comparison}) also demonstrates the effectiveness of FreGrad in reconstructing accurate frequency distributions. In addition, FreGrad takes significant advantage of fast training time. It requires 46 GPU hours to converge, 3.7 times faster than that of PriorGrad with 170 GPU hours.

\begin{table}[ht]
    \centering
        \vspace{-3mm}
        \caption{Ablation study for FreGrad components.}
        \label{tab:ablation_results}
        \resizebox{\columnwidth}{!}{
        \begin{tabular}{lccc}
        \toprule
        & CMOS $\uparrow$ & RMSE$_{f_0}$ $\downarrow$ & RTF on GPU $\downarrow$\\ \midrule
        \textbf{FreGrad}                   &~~~$0.00$  & ${38.73}$ & $0.29$   \\ \midrule
        w/o {Freq-DConv}                 &    $-1.34$    & $39.05$ & ${0.18}$ \\ 
        w/o {separate prior}    & $-0.26$   & $38.91$ & $0.29$ \\ 
        w/o {zero SNR}     & $-0.69$    & $39.17$ & $0.29$   \\ 
        w/o {$\mathcal{L}_{mag}$}   &$-0.68$   & $39.82$ & $0.29$     \\
        \bottomrule
        \end{tabular}
        }
        \vspace{-4mm}
\end{table}

\subsection{Ablation Study on Proposed Components}
\label{subsec:ablation}
To verify the effectiveness of each FreGrad component, we conduct ablation studies by using comparative MOS (CMOS), RMSE$_{f0}$, and RTF. In CMOS test, raters are asked to compare the quality of audio samples from two systems from $-3$ to $+3$.
As can be shown in \Tref{tab:ablation_results}, each component independently contributes to enhancing the synthetic quality of FreGrad. Especially, the utilization of Freq-DConv substantially elevates the quality with a slight trade-off in inference speed, where the increased RTF still surpasses those of existing approaches.
The generation qualities show relatively small but noticeable degradations when the proposed separate prior and zero SNR techniques are not applied.
The absence of $\mathcal{L}_{mag}$ results in the worst performance in terms of RMSE$_{f0}$, which indicates that $\mathcal{L}_{mag}$ gives effective frequency-aware feedback.

\section{Conclusion}

We proposed FreGrad, a diffusion-based lightweight and fast vocoder. FreGrad operates on 
a simple and concise wavelet feature space by adopting a lossless decomposition method.
Despite the small computational overhead, FreGrad can preserve its synthetic quality with the aid of Freq-DConv and the bag of tricks, which is designed specifically for diffusion-based vocoders.
Extensive experiments demonstrate that FreGrad significantly improves model efficiency without degrading the output quality. 
Moreover, we verify the effectiveness of each FreGrad component by ablation studies.
The efficacy of FreGrad enables the production of human-like audio even on edge devices with limited computational resources.

\vfill\pagebreak

\bibliographystyle{IEEEbib}
\bibliography{refs}

\end{document}